\def\mearth{{\rm\,M_\oplus}}

\documentclass{iau}
\usepackage{graphicx}

\title[The Grand Tack: a critical review] %% give here short title %%
{The Grand Tack model: a critical review}

\author[Sean N. Raymond \& Alessandro Morbidelli]   %% give here short author list %%
{Sean N. Raymond$^1$
 \and Alessandro Morbidelli$^2$}
\affiliation{$^1$Laboratoire d'Astrophysique de Bordeaux, CNRS and Universit{\'e} de Bordeaux, \\ UMR 5804, F-33270 Floirac, France. \\ email: {\tt rayray.sean@gmail.com} \\[\affilskip]
$^2$Observatoire de la Cote d'Azur, Laboratoire Lagrange,\\ Bd. de l'Observatoire, B.P. 4229, F-06304 Nice Cedex 4, France. \\email: {\tt morby@oca.eu}}

\pubyear{2014}
\volume{310}  %% insert here IAU Symposium No.
\pagerange{xxx--xxx}
\jname{Complex Planetary Systems}
\editors{Z. Knezevic \& A. Lema\^itre}
\begin{document}

\maketitle

\begin{abstract}
The ``Grand Tack'' model proposes that the inner Solar System was sculpted by the giant planets' orbital migration in the gaseous protoplanetary disk.  Jupiter first migrated inward then Jupiter and Saturn migrated back outward together.  If Jupiter's turnaround or ``tack'' point was at $\sim 1.5$~AU the inner disk of terrestrial building blocks would have been truncated at $\sim 1$~AU, naturally producing the terrestrial planets' masses and spacing.  During the gas giants' migration the asteroid belt is severely depleted but repopulated by distinct planetesimal reservoirs that can be associated with the present-day S and C types. The giant planets' orbits are consistent with the later evolution of the outer Solar System. 

Here we confront common criticisms of the Grand Tack model.  We show that some uncertainties remain regarding the Tack mechanism itself; the most critical unknown is the timing and rate of gas accretion onto Saturn and Jupiter.  Current isotopic and compositional measurements of Solar System bodies -- including the D/H ratios of Saturn's satellites -- do not refute the model.  We discuss how alternate models for the formation of the terrestrial planets each suffer from an internal inconsistency and/or place a strong and very specific requirement on the properties of the protoplanetary disk.  

We conclude that the Grand Tack model remains viable and consistent with our current understanding of planet formation.  Nonetheless, we encourage additional tests of the Grand Tack as well as the construction of alternate models.  

\end{abstract}

\firstsection % if your document starts with a section,
\section{The Grand Tack model}

Until recently, models of terrestrial planet formation suffered from a debilitating problem: they could not form Mars.  Rather, simulations systematically produced planets at Mars' location that were 5-10 times more massive than the real one.  This issue is commonly referred to as the {\it small Mars} problem (Wetherill 1991; Chambers 2001; Raymond et al 2009).

There have long existed solutions to the small Mars problem but none that appeared reasonable (section 3; see also recent reviews by Morbidelli et al 2012 and Raymond et al 2014).  For instance, the terrestrial planets are easily reproduced if the initial conditions for planet formation consisted of just a narrow annulus of large planetary embryos extending from 0.7 to 1 AU (Hansen 2009; see also Morishima et al 2008).  Earth and Venus accreted within the annulus but Mars and Mercury were scattered beyond the edges and effectively starved.  This produced a large Earth and Venus and a smaller Mercury and Mars. The problem is in justifying these initial conditions, especially the sharp outer edge of the annulus.  Given that protoplanetary disks are extended objects and that additional planets exist beyond Mars' orbit, why should terrestrial embryos only exist out to 1 AU?

The Grand Tack model (Walsh et al 2011)\footnote{For a layman-level introduction to the Grand Tack, see http://planetplanet.net/2013/08/02/the-grand-tack/. } proposes that the giant planets are responsible for Mars' small mass.  The model proposes the following train of reasoning, illustrated in Figure~1.  Jupiter was the first gas giant to form.  It accreted gas onto a $\sim10 \mearth$ core (e.g., Pollack et al 1996), carved an annular gap in the disk and migrated inward on disk's local viscous timescale (e.g., Lin \& Papaloizou 1986).  Saturn grew concurrently with Jupiter but farther out and somewhat slower.  Saturn accreted gas and also started to migrate inward.  Saturn's migration was faster than Jupiter's (Masset \& Papaloizou 2003).  Saturn caught up to Jupiter and became trapped in its exterior 2:3 mean motion resonance (Masset \& Snellgrove 2001; Morbidelli \& Crida 2007; Pierens \& Nelson 2008).  This shifted the balance of disk torques acting on the planets' orbits, and Jupiter and Saturn then migrated {\it outward} together (Masset \& Snellgrove 2001; Morbidelli \& Crida 2007; Pierens \& Raymond 2011; D'Angelo \& Marzari 2012).  Outward migration slowed and eventually stopped as the disk dissipated, stranding the gas giants on still-resonant orbits close to their current ones.  This configuration is consistent with a much later instability envisioned by the Nice model (Morbidelli et al 2007; Levison et al 2011; Nesvorny \& Morbidelli 2012).

\begin{figure}[t]
% \vspace*{-2.0 cm}
\begin{center}
 \includegraphics[width=3.4in]{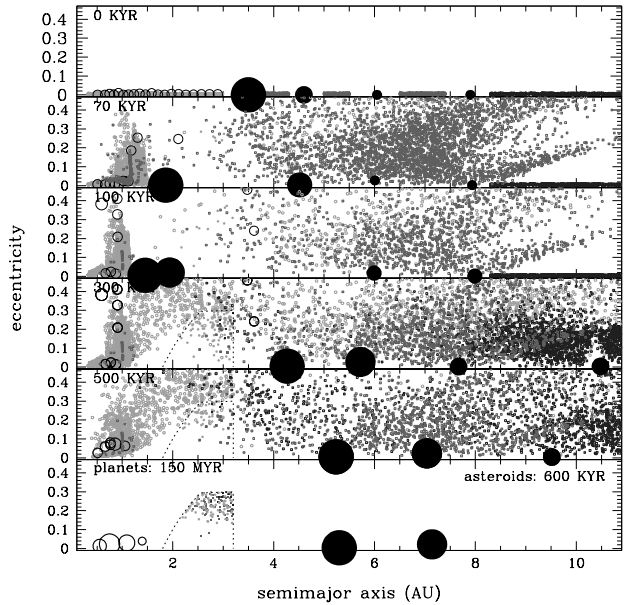} 
% \vspace*{-1.0 cm}
 \caption{Snapshots in time of the evolution of the inner Solar System in the Grand Tack model.  The large black dots represent the giant planets, the open circles represent terrestrial embryos, and the small symbols are planetesimals.  Planetesimals originating interior to Jupiter's orbit are light grey; planetesimals initially between the giant planets are darker grey; and those beyond Neptune are even darker grey.  The dashed curves represent the approximate boundaries of the present-day asteroid belt.  Adapted from Walsh et al (2011; this version from Raymond et al 2014).  }
   \label{fig:evol}
\end{center}
\end{figure}

The inner Solar System was sculpted by the giant planets' migration.  As Jupiter migrated inward it shepherded most of the material interior to its orbit inward (see, e.g., Mandell et al 2007), compressing the disk of embryos.  Jupiter also truncated the inner disk at its closest approach to the Sun.  To truncate the disk of embryos at 1 AU requires Jupiter to be located at $\sim1.5$~AU.  This constrains the ``tack point'', where Jupiter's migration changed direction, which otherwise cannot be estimated {\it a priori}.  As Jupiter and Saturn migrated outward their influence on the inner disk waned.  The giant planets encountered planetesimals from a number of source regions: inner-disk planetesimals scattered outward during the giant planets' inward migration; planetesimals that originated in between the giant planets; and planetesimals that accreted beyond the giant planets, beyond roughly 10-12 AU.  Jupiter and Saturn ejected the majority of these planetesimals but some survived on stable orbits in the asteroid belt.

The Grand Tack reproduces a number of aspects of the inner Solar System.  The disk of embryos that was truncated at 1 AU forms Mars analogs with the correct mass (Walsh et al 2011; O'Brien et al 2014; Jacobson \& Morbidelli 2014).  The asteroid belt is strongly depleted during the giant planets' migration, consistent with the low density of the present-day belt (which contains $\lesssim 10^{-3} \mearth$).  Grand Tack simulations show that the planetesimals implanted during the giant planets' migration are consistent with the observed structure of the present-day belt (Gradie \& Tedesco 1982; Demeo \& Carry 2014).  The inner belt is dominated by particles that originated interior to Jupiter's orbit -- assumed to represent S-types -- and the outer belt is dominated by particles that originated between and beyond the giant planets (Walsh et al 2012).  Finally, the Grand Tack can also explain the origin of Earth's water.  Although Earth forms mainly from the inner disk of embryos that were likely very dry, the inner regions are polluted by water-rich planetesimals.  Polluting planetesimals were scattered inward by the giant planets and overshot the asteroid belt into the inner disk.  Polluting particles outnumber those trapped in the asteroid belt by roughly an order of magnitude and naturally provide Earth with its current water budget (Walsh et al 2011; O'Brien et al 2014).  These particles come from the same parent population as those implanted into the outer asteroid belt and should therefore have C-type compositions and be consistent with the chemical signature of Earth's water (Morbidelli et al 2000; Marty \& Yokochi 2006).  

A number of questions remain for the Grand Tack model.  Is Jupiter's inward-then-outward migration plausible?  Is the large-scale implantation of the asteroid belt consistent with the properties known for Solar System objects?  Do other, simpler models exist to explain the origin of the inner Solar System?  Here we address these questions.  We start by confronting specific issues with the model (section 2).  We then explore the validity of alternate models (section 3).  We conclude in section 4.

\section{Criticisms of the Grand Tack}

\subsection{Dynamics of the Tack mechanism}

The Grand Tack is built on the outward migration of Jupiter and Saturn.  This is the opposite of standard type II migration, which is generally directed inward (Lin \& Papaloizou 1986; Ward 1997; Kley \& Nelson 2012).  For the gas giants to migrate outward they must satisfy two criteria (Masset \& Snellgrove 2001; Morbidelli \& Crida 2007).  First, the planets must orbit close enough to each other that the two planets' annular gaps in the gaseous disk overlap.  The most common orbital configuration is the mutual 3:2 resonance (but other configurations -- such as the 2:1 resonance -- are possible; Pierens et al 2014).  Second, the Jupiter-to-Saturn mass ratio must be between roughly 2 and 4.  

The timescale for inward migration is roughly proportional to the disk's local surface density.  The timescale for accretion is also roughly proportional to the surface density.  Naively, migrating a given distance should therefore increase a planet's mass by a given amount.  If Jupiter and Saturn's cores formed in the same region of the Solar System, how then could Saturn have caught up with Jupiter without growing to a Jupiter-mass?

One possible solution is that the properties of planet migration change as the protoplanetary disk evolves.  Type-I migration -- the regime relevant for low-mass planets and giant planets' cores -- is sensitive to the disk's thermodynamic properties (Masset \& Casoli 2010; Paardekooper et al 2011; Lega et al 2014).  In time, gaseous protoplanetary disks viscously evolve and dissipate (see reviews by Armitage 2011, Alexander et al 2014).  Bitsch et al (2014) used migration maps for planets embedded in disks with a range of properties to show that rapid inward migration is more strongly inhibited in younger disks than in older ones.  Jupiter's inward migration was slowed until it became massive enough to open a gap in the disk and provoke a transition to the type-II migration regime.  Saturn's core, however, could migrate rapidly inward at a much lower mass, i.e. much earlier in its formation history.  This may well have allowed Saturn to catch up with Jupiter and produce the configuration needed for outward migration.  In simple terms, the above argument fails because migration does not scale linearly with the disk's mass.  

Jupiter and Saturn may thus achieve the required configuration for outward migration, but can said configuration be maintained throughout migration?  If one planet accretes much faster than the other then their mass ratio may be driven outside of the range needed for outward migration.  Unfortunately, our understanding of gas accretion onto giant planet cores remains incomplete.  Giant planets accrete gas first by the slow collapse of quasi-hydrostatic envelopes (Ikoma et al 2000, Machida et al 2010) and later by viscous transport through a circumplanetary disk (Ayliffe \& Bate 2009; Uribe et al 2013).  Some models suggest that circumplanetary disks may have very low viscosities and thus act as bottlenecks for giant planet growth (Fujii et al 2011; Rivier et al 2012).  Yet there remain additional sources of accretion, for example from polar inflows (Morbidelli et al 2014; Szulagi et al 2014).  Hydrodynamical simulations of planet migration do not have the requisite resolution to realistically include gas accretion, yet these two are intimately coupled in the Grand Tack model.  This is a key uncertainty for the Grand Tack: it is unclear whether long-term outward migration of Jupiter and Saturn is possible given the stringent mass ratio requirement.  Yet it is a fact that even today Jupiter and Saturn have the appropriate mass ratio for outward migration.

\subsection{Compositions and isotopic ratios of small bodies}

The Grand Tack model proposes that the present-day asteroid belt was implanted from planetesimals that accreted across the Solar System (Walsh et al 2011, 2012).  After being scattered outward-then-inward, the S-types in the inner belt remain close to their original orbital radii.  The C-types, however, were scattered inward and trapped in the main belt from between and beyond the giant planets (see Fig. 1).  C-types are thought to be represented by carbonaceous chondrite meteorites.  Although there is a spread in values, the D/H ratios of carbonaceous chondrites are a good match to Earth's water (Marty \& Yokochi 2006; Alexander et al 2012).  The D/H ratios of bodies originating in the outer Solar System are more uncertain.  The D/H ratios of nearly-isotropic comets thought to originate in the Oort cloud are roughly twice as high as Earth's (e.g., Bockelee-Morvan et al 2012).  Classically, the Oort cloud comets are expected to have formed from the giant planet region, but recently Brasser and Morbidelli (2013) argued in favor of a trans-Neptunian origin. The D/H ratio of Saturn's moon Enceladus is also roughly twice Earth's (Waite et al 2009).  

Alexander et al (2012) used these data to argue against large-scale implantation of C-types from the outer Solar System.  They proposed that Enceladus' and Oort cloud comets' elevated D/H ratios are characteristic of planetesimals formed in the giant planet region. This would mean that planetesimals formed beyond Saturn could not be precursors of C-type asteroids.  However, the D/H ratio of Saturn's largest moon, Titan, has been measured to be Earth-like (Coustenis et al 2008; Abbas et al 2010; Nixon et al 2012).  Given its much larger mass, Titan's D/H -- assuming it represents the bulk source of water -- is much more likely than Enceladus' to be representative of the D/H of locally-grown planetesimals.  In addition, the disparity in D/H between the two satellites calls into question the very notion of using moons' compositions to constrain dynamical models.  Of course, there remain several uncertainties.  While Enceladus' D/H ratio was measured in water, Titan's was measured in methane and acetylene.  The D/H ratio of Titan's water is not certain.  Strong fractionation may produce different D/H ratios for different species.  In the case of comet Hale-Bopp, the D/H ratio measured in HCN was 7 times larger than that measured in water (Meier et al 1998).  What counts is the D/H ratio of the dominant reservoir of H.  This remains unknown for Saturn's moons.  

Alexander et al also argued that a correlation between D/H and C/H in meteorites shows isotopic exchange between the pristine ice and organic matter within the parent bodies of carbonaceous chondrites.  This would suggest that the original water reservoir had an even lower D/H than Earth, Titan or any comet, again making carbonaceous asteroids distinct from bodies formed in the giant planet region.  However, such a reservoir of pristine ice has never been observed; the fact that Earth's water and other volatiles are in chondritic proportion means that carbonaceous chondrites -- wherever they formed -- reach their current bulk D/H ratios very quickly, before delivering volatiles to Earth.  The same could have happened to comets and satellites.  

Another argument for a distinction between carbonaceous asteroids and comets is that even the comets with a chondritic D/H ratio (e.g. Hartley 2; Meech et al., 2011) have a non-chondritic $^{15}$N/$^{14}$N ratio. Titan has a cometary $^{15}$N/$^{14}$N as well (Mandt et al., 2014). Here, again, a few caveats are in order. First, it is difficult to relate a satellite composition, born from a circum-planetary disk with its own thermal and chemical evolution, to the composition of bodies born at the same solar distance but on heliocentric orbits.  Second, it is unclear whether any comets for which isotope ratios have been measured originate from the giant planet region, as opposed to the trans-Neptunian region (see Brasser \& Morbidelli 2013).

A similarity between carbonaceous chondrites and comets has been proposed from the analysis of micro-meteorites.  The isotopic ratios of most micro-meteorites -- $\sim 100 \mu$m particles collected in Antarctic ice -- are chondritic (with the exception of the ultra-carbonaceous particles, which constitute a small minority of micrometeorites; Duprat et al 2010).  However, dynamical models show that most of the dust accreted by Earth should be cometary, even taking into account the entry velocity bias (Nesvorny et al., 2010; Rowan-Robinson \& May 2013).  For the case of the CI meteorite Orgueil, Gounelle et al (2006) used an orbital reconstruction to argue for a cometary origin.  These factors suggest that the rocky components of comets and carbonaceous asteroids are very similar -- perhaps indistinguishable -- in their bulk- and isotopic compositions.  

%Third, the D/H ratios of two Jupiter-family comets are a good match to Earth's water (Hartogh et al 2011; Lis et al 2013).  These comets are thought to have originated in the Kuiper belt (e.g., Levison \& Duncan 1994).  This is farther out than the presumed source region for Oort cloud comets (e.g., Dones et al 2004; but see Brasser \& Morbidelli 2013).  The primordial radial D/H gradient may not have been monotonic. Finally, although the D/H ratios of Jupiter-family comets match Earth's, their $^15$N/$^14$N ratios are roughly twice as high (similar to Oort cloud comets'; Hartogh et al 2011).  Of course, as discussed previously, it is unclear how well the observed isotopic ratios represent those for the bulk reservoirs of these species.  In addition, the statistics for of isotope ratios for comets are very sparse.  

It has been argued that if the parent bodies of carbonaceous chondrites accreted among the giant planets they should be $\sim 50\%$ water rather than the $\sim10\%$ inferred from meteorites (Krot 2014). However, a body's original water content cannot easily be estimated from its aqueous alteration.  The carbonaceous parent bodies may very well have been more water-rich than the alteration seems to imply.  In addition, large main belt comets Themis (Campins et al 2010; Rivkin \& Emery 2010) and Ceres (Kuppers et al 2014) appear to contain far more water than carbonaceous chondrites.  Meteorites may simple represent rocky fragments of bodies that were far wetter/icier.

\section{Alternate models}

%We now discuss other models for solving the small Mars problem and forming the terrestrial planets.  

It is often assumed that giant and terrestrial planet formation can be considered separate phases.  Most simulations of terrestrial accretion start from fully-formed giant planets that perturb the building blocks of terrestrial planets.  Following Raymond et al (2014), we refer to these as {\it classical} models.  Of course, there are nonetheless free parameters, in particular the distribution of planetesimals and planetary embryos and the giant planets' orbits.

\subsection{Effect of the giant planets' orbits}

In order to fit in a self-consistent model of Solar System formation, the giant planets' orbits must be chosen carefully.  For instance, it is inconsistent for Jupiter and Saturn to be on their current orbits during terrestrial accretion.  Scattering of embryos by giant planets during accretion systematically decreases the giant planets' orbital eccentricities (e.g., Raymond 2006).  If they were at their current orbital radii, Jupiter and Saturn must therefore have had higher eccentricities at early times.

The simplest version of the classical model assumes that Jupiter and Saturn's early orbits can be represented by the initial conditions for the instability presumed to have caused the late heavy bombardment (the Nice model; Gomes et al 2005).  A typical configuration places Jupiter and Saturn in 3:2 resonance with Jupiter at $\sim 5.4$~AU and low eccentricities for both planets (Morbidelli et al 2007).  Simulations of terrestrial accretion with Jupiter and Saturn on these orbits completely fail to reproduce the terrestrial planets (Raymond et al 2006, 2009; Morishima et al 2010; see also Fischer \& Ciesla 2014).  Mars analogs are far too large (hence the small Mars problem) and Mars-sized embryos are systematically stranded in the asteroid belt.  

Some classical model simulations can reproduce Mars' small size.  To date, the most successful is the {\it EEJS} -- for `Extra Eccentric Jupiter and Saturn' -- model of Raymond et al (2009) and Morishima et al (2010).  In this model the gas giants' primordial eccentricities were somewhat larger than their current values: 0.07-0.1 rather than $\sim 0.05$.  The $\nu_6$ secular resonance at 2.1 AU -- which marks the inner edge of the main asteroid belt -- is much stronger with this orbital configuration of the giant planets.  Particles that enter the $\nu_6$ quickly have their eccentricities pumped to high values and are lost from the system (Gladman et al 1997).  Given the orbital `jostling' caused by interactions between embryos, the $\nu_6$ acts to drain material out of Mars' feeding zone and restrict its mass.  

A similar model was proposed by Nagasawa et al (2005) and Thommes et al (2008).  In their model the $\nu_5$ secular resonance swept inward through the asteroid belt as the protoplanetary disk dissipated, asymptotically reaching its current location at $\sim0.7$~AU (Ward 1981).  Strong dynamical excitation from the sweeping secular resonance efficiently cleared out the Mars region without depleting the growing Earth or Venus.

Despite the apparent success of the EEJS and secular resonance sweeping models they are hard to put in context.  The sweeping secular resonance model (Thommes et al 2008) requires the presence of a gas disk, yet it is not consistent with accepted theory for planet-disk interactions as it requires Jupiter and Saturn to maintain their current orbits in the presence of the disk.  The disk should realistically damp out any residual eccentricity and drive the planets into a resonant configuration (e.g., Morbidelli et al 2007; Kley \& Nelson 2012).  The EEJS model suffers from a similar problem, as no model has ever been proposed to explain such large eccentricities of the giant planets either late in the gaseous disk phase or immediately after.  One might imagine that a Nice model-like instability could have occurred immediately after the dispersal of the disk; however, such an instability can only reasonably excite Jupiter and Saturn's eccentricities to their current values, not to values twice as large.  In addition, there would necessarily be a prolonged phase of planetesimal scattering associated with such an instability, which would act to spread the orbits of Jupiter and Saturn and effectively change the position of the $\nu_6$ resonance.  Finally, if the EEJS model represents the correct initial orbital configuration, then no later orbital migration of the giant planets is allowed.  Of course, late migration is needed to explain the orbital distribution of the Kuiper belt (Malhotra 1995; Levison et al 2008; Dawson \& Murray-Clay 2012) and the late heavy bombardment (Gomes et al 2005; Morbidelli et al 2007; Levison et al 2011).

\subsection{Effect of the disk properties}

A depletion of mass in the Mars region is clearly required to solve the small Mars problem.  Perhaps small bodies simply migrated inward and cleared the Mars region (Kobayashi \& Dauphas 2013).  Inward migration has been invoked to explain systems of super-Earths observed on close-in orbits around other stars (Terquem \& Papaloizou 2007; Cossou et al 2014).  While no super-Earths exist in the Solar System, could a similar mechanism have been at play?  Indeed, large-scale migration could explain the asteroid belt's strong mass depletion.  However, if embryos migrated inward {\it en masse} from the asteroid belt, the eccentricities and inclinations of planetesimals that were left behind should be very low due to aerodynamic drag from the gaseous disk.  This conflicts with the belt's broad eccentricity and inclination distributions.  Apart from the Grand Tack, the only known mechanism capable of producing the observed distributions is scattering by embryos within the belt {\it after} the dissipation of the disk (Petit et al 2001, Chambers \& Wetherill 2001; O'Brien et al 2007).  Of course, invoking this mechanism invalidates our assumption that embryos migrated inward away from the belt.  
%To date, the Grand Tack is the only model that can simultaneously produce a small Mars and the correct asteroid belt structure.  

Jin et al (2008) proposed that the viscosity structure of the protoplanetary disk could produce the required mass depletion.  In their model the inner and outer parts of the disk are MRI-active and therefore high-viscosity as they are ionized by the irradiation from the central star and external OB stars.  However, the middle region has a low viscosity and is essentially a dead zone.  A deficit in the disk's surface density is created at the inner edge of the dead zone, with a viscous inner disk and an inviscid outer disk.  For certain disk parameters, Jin et al found that this deficit can be located close to Mars' orbit.  
Izidoro et al (2014) tested the effect of such a deficit on the accretion of the terrestrial planets.  They found that in some cases a small Mars could indeed form, and water could also be delivered to Earth from more distant C-type material (as in the classical model; Morbidelli et al 2000; Raymond et al 2007).  Their successful simulations reduced the surface density in the Mars region by a factor of 4 and placed the giant planets at their current orbital radii and with their current orbital eccentricities.  As discussed above, it is not strictly self-consistent for the giant planets to be on their current orbits during this epoch.  This model therefore places not one but two stringent requirements on Solar System formation.  First, the disk must have a carefully-specified viscosity structure.  And second, no late migration or eccentricity damping of the giant planets is allowed.  As discussed above, this second requirement flies in the face of current thinking about the evolution of the outer Solar System.

\section{Conclusions}

This short review served to address common criticisms of the Grand Tack model.  Given our current understanding the Grand Tack is self-consistent and provides a reasonable solution to the small Mars problem.  There is one clear loose end related to the coupling between gas accretion onto the giant planets and outward migration (section 2.1).  This is important because a Jupiter-to-Saturn mass ratio of 2-4 is required for outward migration to occur.  We showed that the compositions and isotopic ratios of known Solar System bodies cannot currently be used to refute the Grand Tack model (section 2.2).  
%Given that two of Saturn's moons have D/H ratios that differ by a factor of two it is unclear how to constrain formation models with such measurements.

We found that each competing model that form a small Mars has strong limitations (section 3).  The EEJS model (section 3.1) is inconsistent with models for the evolution of the outer Solar System.  The sweeping secular resonance model (section 3.1) is not consistent with the well-developed theory of planet-disk interactions.  Finally, the model proposing a mass deficit in the disk in the Mars region (section 3.2) both requires a fine-tuned disk structure and does not allow late evolution of the outer Solar System.  

Nonetheless, we hope and fully expect that new models will provide alternate pathways to solving the small Mars problem.  

%For instance, it is possible that that planetesimals may only form in specific discrete regions within the disk that have the right conditions.  

\vskip .2in
\noindent {\bf Acknowledgments.} 
This work was funded by the Agence Nationale pour la Recherche via grant ANR-13-BS05-0003-002 (project {\em MOJO}).

\end{document}